# Anatomy of interfacial spin-orbit coupling in Co/Pd multilayers using X-ray magnetic circular dichroism and first-principles calculations

Jun Okabayashi,[1*] Yoshio Miura,[2] and Hiro Munekata[3]

[1]Research Center for Spectrochemistry, The University of Tokyo, Bunkyo-ku, Tokyo 113-0033, Japan

[2]Research Center for Magnetic and Spintronic Materials, National Institute for Materials Science (NIMS), Tsukuba 305-0047, Japan

[3]Laboratory for Future Interdisciplinary Research Science and Technology, Tokyo Institute of Technology, Yokohama 226-8503, Japan

*To whom correspondence should be addressed, e-mail: jun@chem.s.u-tokyo.ac.jp

(24 April 2018)

Element-specific orbital magnetic moments and their anisotropies in perpendicularly magnetised Co/Pd multilayers are investigated using Co $L$-edge and Pd $M$-edge angle-dependent x-ray magnetic circular dichroism. We show that the orbital magnetic moments in Co are anisotropic, whereas those in Pd are isotropic. The first-principles density-functional-theory calculations also suggest that the Co/Pd interfacial orbital magnetic moments in Co are anisotropic and contribute to the perpendicular magnetic anisotropy (PMA), and that the isotropic ones in Pd manipulates the Co orbitals at the interface through proximity effects. Orbital-resolved anatomy of Co/Pd interfaces reveals that the orbital moment anisotropy in Co and spin-flipped transition related to the magnetic dipoles in Pd are essential for the appearance of PMA.

From magnetic-pin or antiferromagnetic-ferromagnetic bilayers in the spin valves developed in late 80's to the most recent perpendicular-anisotropy CoFeB-MgO layers in magnetic tunnel junctions [1], it is one of the central research subjects in spintronics to pursue ways to control long-range spin ordering and resultant magnetization vectors starting from the interface of constituent nanostructures. Among various interface combinations, the interplay between 3$d$ transition metals (TMs) and 4$d$ or 5$d$ TMs has been considered to offer opportunity of studying an interface-driven magnetic anisotropy with both spin and orbital degrees of freedom; namely the interplay between the two, the spin-orbit interaction through interfacial chemical bonding. Ultrathin Co/Pd multilayers are one of such representative artificial nanomaterials that exhibit *interface* perpendicular magnetic anisotropy (PMA), and the development of artificially synthesised PMA has led researchers to the expectation of ultra-high density recording media [2,3]. Since then, extensive efforts have been made for studying electronic and spin structures of the interfaces of ultra-thin magnetic multilayers [4-12] and nanostructures [13-16]. Studies on Co atoms performed using x-ray magnetic circular dichroism (XMCD) have suggested the enhancement of orbital magnetic moments at the interfacial Co that is adjacent to Pd or Pt. It has been reported that the PMA emerges due to the cooperative effects between spin moments in 3$d$ TMs and large spin-orbit interactions $\xi LS$ in the non-magnetic 4$d$ or 5$d$ TMs, where $\xi$ is the spin-orbit coupling constants, $L$ is orbital angular momentum, and $S$ is the spin angular momentum [4-16]. The Co/Pd interfaces and multilayers have also been employed to demonstrate the photo-induced precession of magnetisation [17, 18], the creation of skyrmions using the interfacial Dzyaloshinskii-Moriya interaction [19], and magnetisation reversal using the spin-orbit torque phenomena [20]. Despite the abovementioned careful studies and interesting trials with Co/Pd interfaces, the interfacial PMA, in particular the mechanism of *anisotropic* orbital magnetic moments, has not been fully understood for both Co and Pd sites. Bruno and van der Laan theoretically proposed an orbital moment anisotropy in 3$d$ TMs within the second-order perturbation of the spin-orbit interaction (the weak coupling) for more than half-occupied electrons [21, 22]. However, in the case of strong spin-orbit coupling in 4$d$ or 5$d$ TMs, the validity of this perturbative formula has been debated [23]. In order to study the mechanisms of PMA in Co/Pd multilayers, the contributions of orbital magnetic moments in each element should be explicitly considered.

However, it is challenging to study the anisotropy of the orbital magnetic moments of both Co and Pd elements using one specific experiment, due to the challenges in detection of the induced magnetic moments, of Pd in particular. In this study, we aim to overcome the above experimental challenges using the x-ray photon energy region that is common for both Co and Pd. In other words, a single experiment is performed using the Pd $M_{2,3}$-edge absorption at ~ 530 eV, and Co $L_{2,3}$-edge absorption at ~ 770 eV taken by electron yield mode, instead of using different energy ranges for the Pd $L$-edges within 3.2 keV taken by fluorescence yield mode [9, 10, 12-15]. The optical transitions from 2$p$ to 3$d$ states in 3$d$ TMs, and those from 3$p$ to 4$d$ states in 4$d$ TMs have the same transition probabilities. Therefore, the same transition processes could occur, even though the cross-section for the $M$-edges is smaller than that for $L$-edges. Furthermore, we adopt the angle-dependent XMCD that is a powerful experimental technique for studying anisotropic orbital magnetic moments. Taking into account that the X-ray absorption spectrum (XAS) attributed to the Pd $M_{2,3}$-edges (3$p$ to 4$d$ transition) overlaps with that of the O $K$-edge absorption region [6, 11, 24], the surface oxide components have to be carefully removed, or the sample has to be prepared *in-situ* in a detection chamber, in order to detect the Pd $M$-edge absorption signals. In this study, we have chosen the Ar-ion sputtering methods for surface cleaning. Experiments thus implemented lead us to the finding that the orbital magnetic moments in Co are anisotropic, whereas those in Pd are isotropic.

Motivated by the experimental finding, theoretical examination incorporating the strong spin-orbit coupling at the interfacial atoms is carried out. The magnetic dipole contribution, which is derived from spin-flipped transitions beyond the orbital magnetic moments, through the quadrupole-like interactions between dipoles is assessed quantitatively with the first-principles density functional theory (DFT) calculations. We focus on the anisotropy of orbital magnetic moments at the Co/Pd interfaces using the DFT calculations [25-28], which provide the element- and layer-resolved contributions that reveal the mechanism of PMA. Consequently, we come to the conclusion that the orbital moment anisotropy in Co and spin-flipped transition related to the magnetic dipoles in Pd are essential for the appearance of PMA.

## Results

### XMCD study.

Two types of ultrathin [Co/Pd]$_5$ multilayered structures were studied: Co (0.69 nm)/Pd (1.62 nm) that have PMA (Sample A), and Co (1.03 nm)/Pd (1.62 nm) that have an in-plane anisotropy (Sample B) as illustrated in Fig. 1. These samples were originally grown to study the photo-induced precession of magnetisation at a low power excitation [18]. The 0.69-nm-thick Co and 1.62-nm-thick Pd in the Sample A correspond to four monolayers (MLs) and eight MLs thickness, respectively. Cross-sectional transmission electron microscopy images have shown ⟨111⟩-oriented layered metallurgical structures with a good interface abruptness between Co and Pd [18]. Element-specific magnetisation hysteresis curves in normal incidence (NI) and oblique or grazing incidence (GI) geometries, taken at magic angle of 54.7° ($\cos^2\theta = 1/3$) from the sample surface normal, for Co are shown in Fig. 1. The photon energies are fixed at 778 eV for Co $L_3$-edge. In the case of GI set up in Sample A, the contributions from both PMA and in-plane anisotropy are included almost equally, which can be helpful to correct the in-plane orbital magnetic moment as discussed later. On the other hand, the in-plane anisotropy that is attributed primarily to the shape anisotropy is observed for the Sample B, as expected. These results are consistent with the magnetisation measurements reported in Ref. 18.

Figure 2 shows the Co $L$-edge XAS and angular-dependent XMCD for the Samples A and B that exhibit PMA and in-plane magnetic anisotropy, respectively, recorded at a temperature of 10 K. The XMCD spectra for the NI and GI geometries, and the integrals of the Co $L_{2,3}$ absorption edges XMCD spectra, are shown in Figs. 2b and 2c, respectively. There are clear differences in XAS spectra, depending on the relative helicities of the incident beam. As the XAS spectra obtained from the NI and GI configurations are identical, only the XAS spectra in the NI configuration are shown. In Fig. 2b, the XMCD spectra for the NI and GI configurations show a distinct difference in intensity between the $L_3$ edges, while the $L_2$ edges have almost similar intensity profiles. For Sample A, the measured XMCD signal for the Co $L_3$ edge in the NI geometry was larger than that in the GI geometry when the effective field is perpendicular to the film plane. On the other hand, Sample B exhibits similar XMCD line shapes for both NI and GI. The magneto-optical sum rule indicates that the integrated areas of both negative $L_3$ and positive $L_2$ peaks are proportional to the orbital magnetic moments. For Sample A, the residuals of the integrals of both $L_3$ and $L_2$ edges in the XMCD spectra are larger in the NI configuration, compared to the GI configuration, which indicates that large orbital magnetic moments are observed for the NI configuration as shown in Fig. 2c. On the other hand, for Sample B, isotropic orbital moments are suggested, which is in agreement with the in-plane anisotropy character. Assuming the Co hole number is 2.49 [29], we deduce that the orbital moments of the perpendicular and in-plane components for Sample A, $m_{\text{orb}}^{\perp}$ and $m_{\text{orb}}^{\parallel}$, are 0.14 $\mu_B$ and 0.10 $\mu_B$, respectively, using the manner for the estimations within the XMCD sum rules [29]. Note that the setup of $m_{\text{orb}}^{\parallel}$ cannot detect the perfect in-plane contribution and almost half of $m_{\text{orb}}^{\perp}$ and $m_{\text{orb}}^{\parallel}$ are mixed which is proven by the MH curves in Fig. 1. For the anisotropy of the orbital moments ($\Delta m_{\text{orb}}$), we obtain a value of 0.04 $\mu_B$ for Co after the correction of $m_{\text{orb}}^{\parallel}$ value (See Method section). Further, the spin magnetic moment ($m_s$) and magnetic dipole moment ($m_T$) of Co are 1.2 $\mu_B$ and 0.01 $\mu_B$, respectively, with the uncertainties of ±10 %. On the other hand, the Sample B has an $m_{orb}$ of 0.09 $\mu_B$. As the Co layer is thicker than the threshold PMA thickness, the contribution from the in-plane shape anisotropy becomes dominant. These values are listed in Table I by comparing with those estimated from the DFT calculations.

Figures 3 show the XMCD signals of Sample A for the Pd $M$-edges, after the removal of surface contamination. The signals that emerge due to the O $K$-edge absorption are removed by the Ar ion sputtering, hence clear XMCD signals (recorded at 10 K) are observed. These signals are induced by the proximity effects with the Co layers. We note that the intensity scale of the Pd $M$-edge is two-orders of magnitude smaller than that of the Co $L$-edge, due to the difference in photo-ionisation cross-section. The Pd $M_{2,3}$-edge XAS line shapes exhibit satellite structures that appear at 542 eV and 567 eV, with higher photon energies of the main absorption peaks. They are obtained from the transitions from 3$p$ to 5$s$ states [6], and do not contribute to the XMCD signals. The Pd $M$-edge XMCD line shapes in both PMA and in-plane samples are almost identical, which suggests isotropic orbital moments in Pd, within the detection limits. Conventional magneto-optical sum rule analysis, often employed for 3$d$ TMs (from 2$p$ to 3$d$ transition), can be applicable for the 3$p$ to 4$d$ transition in Pd $M$-edge XMCD with the same transition probabilities as that of the 2$p$ to 3$d$ transition. Assuming the hole number of the Pd 4$d$ states is 1.10 [14], for the NI configuration, the estimates of the spin and orbital magnetic moments are 0.25 $\mu_B$ and 0.02 $\mu_B$, respectively, with the error bars of ±20 % because of the estimations of XAS spectral integrals include the ambiguities (See Method section). For both PMA and in-plane samples, in contrast to Co, the Pd XMCD line shapes remain almost unaffected in the angular dependence (within the detection limits). This indicates that the isotropic finite orbital moments in Pd do not directly contribute to the PMA. In the previous papers using Pd $L$-edges, almost quenched orbital magnetic moments are reported [12,13]. These contradicts might be derived from the differences in detection modes or sample structures. In order to further study the effects of the Pd layer thickness, we measured the PMA samples that have a 0.8-nm-thick Pd layer. The results indicate that similar tendencies are observed using the angle-dependent XMCD in Pd *M*-edges.

**First-principles calculation.**

Figure 4a shows the magneto-crystalline anisotropy energy ($E_{MCA}$), demagnetisation energy $2\pi M_S^2$, and $E_{MCA} - 2\pi M_S^2$ as a function of the Co layer thickness ($t_{Co}$), where $M_S$ is the magnetisation of the multilayer. The $2\pi M_S^2$ is defined as the demagnetisation energy of Co(*n*)/Pd(8 ML) that contributes to the in-plane magnetic anisotropy. As shown in Fig. 4a, the $E_{MCA}$ decreases with the increase of $t_{Co}$, while the demagnetisation energy $2\pi M_S^2$ increases with $t_{Co}$. Therefore, the PMA is reduced when $t_{Co}$ increases, and the in-plane anisotropy is preferred when $t_{Co}$ is larger than 1.5 nm, which is qualitatively consistent with the magnetisation characteristics of Samples A and B. The increase of the $E_{MCA}$ with the decrease of $t_{Co}$ can be explained by the strong interfacial perpendicular MCA of the Co/Pd(111) as well as the small contribution to the MCA from the bulk fcc-Co. We consider the (111) interfaces, as illustrated in Fig. 4b. The chemical bonding along the *z*-direction is staggered between Co and Pd, which causes the non-perfect σ bonding at the interface. Therefore, the isotropic distribution of orbital moments in Pd, and the anisotropic orbital moments in the Co layer at the interface promotes the PMA at the Co/Pd interface. On the other hand, it has been reported that the Pd(001) orientation cannot stabilise the PMA at the interface [30,31]. Furthermore, three monolayers are necessary for the periodic stacking in (111) orientation, as shown in Fig. 4b. Therefore, (111) orientation stacks are necessary for the PMA, due to the tuning of the interfacial hybridisation strength.

In order to analyse the origin of the PMA, we computed the orbital moment anisotropy at each atomic site. Spin and orbital magnetic moments in Co and Pd sites are listed in Table I. As shown in Fig. 4c, the orbital moment anisotropy of the Co atoms at the Co(4 ML)/Pd(8 ML) interfacial layer is enhanced and has a value of 0.033 $\mu_B$. We emphasise that the PMA of the Co monolayer that is next to the Pd monolayer can be augmented primarily by the contribution of the interfacial layer. The PMA in a Co monolayer decreases as the distance between this monolayer and the Co/Pd interface increases, which is expected due to the bulk-like Co-Co bonding. On the other hand, we estimate that the orbital moment anisotropy induced in Pd is very small compared with that in Co, even in the Pd layer that is next to Co (Fig. 4c). The induced magnetic moments in the Pd monolayer emerge due to the *d*-orbital hybridisation between neighboring Co 3*d* and Pd 4*d* states at the interface, which qualitatively coincides with the independence of orbital magnetic moment anisotropy in Pd sites deduced from the XMCD.

Figures 5a and 5b represent the contributions of the crystalline magnetic anisotropy on the anisotropy energy at each atomic site. Four types of spin transition processes occur between the occupied and unoccupied states within the second-order perturbation of the spin-orbit interaction. The "up-down" process implies a virtual excitation from an occupied up-spin state to an unoccupied down-spin state in the second-order perturbation. For Co sites, the transition between down-down spin states is dominant, as shown in Fig. 5a. This suggests the conservation of spin states in the transition, which can be explained using the Bruno model assuming a large spin splitting. The Co sites exhibit positive energies in total, which confirms the orbital-moment-driven PMA. In contrast, for Pd, the spin-flipped transitions between up-down and down-up states become dominant due to the small band splitting, hence both spin-preserved and spin-flipped processes occur near the Fermi level. Element- and orbital-resolved density of states (DOS) for the Co 3*d* and Pd 4*d* states are shown in Figs. 5c and 5d, respectively. The DOS of both Co 3*d*($xy$, $x^2$-$y^2$) and 3*d*($yz$, $zx$) orbitals at the interface contribute to the PMA because the 3*d*($xy$, $x^2$-$y^2$) and 3*d*($yz$, $zx$) orbitals become dominant at the Fermi level in the minority-spin state. These states provides large matrix elements of $L_z$, $<3d(x^2-y^2)\downarrow|L_z|3d(xy)\downarrow>$ and $<3d(yz)\downarrow|L_z|3d(zx)\downarrow>$, for the second order perturbation of the spin-orbit interaction, leading to an enhancement of the perpendicular components of the orbital magnetic moments [32]. The DOS of the Co 3*d* states is clearly split, with a spin magnetic moment of 1.89 $\mu_B$. We used the lattice constant of $a = 0.391$ nm for the equilibrium condition. The DOS of Pd 4$d$ states is also split, due to the proximity with the Co layers, as shown in Fig. 5d. For the Pd 4$d$ states, we estimated that the induced spin magnetic moment is 0.311 $\mu_B$. The Pd 4$d$ states exhibit a small splitting at the interfaces, whereas the DOS at the Fermi level is large, which is considered as a Stoner-type ferromagnetism. The matrix elements of <4$d(yz)$↓|$L_z$|4$d(x^2-y^2)$↑> are dominant in Pd sites, which favours to in-plane anisotropy. The spin-conserved transition in Co and spin-flipped transitions in Pd are illustrated in Fig. 5e. The enhancement of orbital moments of Co is explained by the spin conserved transition derived from the band structures. The spin-orbit coupling in heavy-metal elements causes a quadrupole-like formation by the spin-flip transitions, resulting in the magnetic dipole term ($m_T$), however it does not contribute to the anisotropy of the orbital moments. These results explain both the angular dependence of the Co $L$-edge and the Pd $M$-edge XMCD spectral line shapes.

**Discussion**

Considering the results of XMCD and DFT calculations, we discuss the quadrupole-like contribution of the interfacial Pd layer. The spin sum rule includes not only the spin moment $m_s$, but also the magnetic dipole term $m_T$, and reveals the effective spin magnetic moment $m_s^{eff} = m_s + 7m_T$. Here, the $m_T$ can be separated from the angular dependence of XMCD, as the GI configuration cancels out the $m_T$ term for the magic angle geometry of 54.7° ($\cos^2\theta = 1/3$) with respect to the surface normal [5,33]. Our Pd XMCD results indicate that the Pd orbital moments induced at the interface are isotropic. Note that $m_T$ is an order of magnitude smaller than the orbital moments, i.e., 0.01 $\mu_B$ or less, comparable with the detectable limits. The element-specific $E_{MCA}$ that includes the $m_T$ term beyond the Bruno model which is derived from only orbital moment anisotropy $\Delta m_{orb}$ can be expressed theoretically as [21, 22]:

$$E_{MCA} \sim -\frac{1}{4\mu_B}\xi\Delta m_{orb} + \frac{21}{2\mu_B}\frac{\xi^2}{\Delta E_{ex}}\Delta m_T \qquad (1)$$

where $\Delta E_{ex}$ is the exchange splitting between spin-up and spin-down bands. $\Delta m_T$ satisfies the relation of $\Delta m_T = m_T^\perp - m_T^\parallel$ with $m_T^\perp = -2m_T^\parallel$. For Co, we estimate the first and second terms in eq.(1) contributing $10^{-4}$ and $10^{-5}$ eV, respectively, in Co, assuming $\Delta E_{ex\ Co} = 3$ eV, and $\xi_{Co} = 70$ meV, and the orbital moment anisotropy becomes dominant even in the finite $m_T$ value. For Pd, the orbital anisotropy that corresponds to the first term becomes almost zero and the second term becomes dominant in the order of $10^{-4}$ eV because of small $\Delta E_{ex\ Pd} = 200$ meV and large $\xi_{Pd} = 110$ meV, even if the $m_T$ in Pd is as small as 0.01 $\mu_B$. The relatively large spin-orbit coupling constant and small Pd exchange splitting contribute to the appearance of PMA by means of the second term in eq.(1) with quadrupole-like interactions. The second term in Pd is comparable or smaller than the orbital moment anisotropy in Co. Therefore, the first term in eq.(1) of orbital moment anisotropy in Co and the second term related to $m_T$ in Pd contribute dominantly to $E_{MCA}$ through the interfacial proximity effects. Furthermore, the contribution to the first term of eq.(1) in Co is underestimated because Fig. 4c suggests the enhancement of $\Delta m_{orb}$ at the interfacial layer. XMCD detects the signals from all layers, which suppresses the contribution from the interfacial layers. However, the enhancement of the orbital moment anisotropy can be concluded qualitatively.

Finally, we discuss the hybridization and spin-orbit coupling at the Co/Pd interfaces. Vogel *et al.* reported that 4-ML Pd layers are polarised through the interfacial proximity effects [12]. On the other hand, our findings indicate that the orbital moments in Pd are isotropic. This suggests that the magnetic dipole transitions can be essential, due to the mixing of spin-up and spin-down states in the Pd 4$d$ states, even though the $m_T$ values are much smaller than $m_{orb}$. As another physical origin of PMA, the facts that the radii of the Pd 4$d$ orbitals are larger than those of Co 3$d$ are also important [34], which results in the decrease of the anisotropy of the orbital moments and the increase of lattice strains. These effects couple at the interfaces, which enhances the PMA in Co/Pd multilayers. For relatively thick Co layers, the shape anisotropy in Co governs and suppress the effects of PMA at the interface, resulting in the in-plane anisotropy. The origin of PMA at the Co/Pd interface can be explained by the interfacial Co orbital moment anisotropy and spin-flipped processes at the Pd sites through the strong hybridization at Co/Pd interface, as illustrated by the magnetic dipoles in Fig. 5e.

In summary, we have investigated the origin of PMA at Co/Pd interfaces using the angle-dependent XMCD and DFT calculations. The Co $3d$ orbital states are anisotropic, while the Pd $4d$ orbital states are isotropic. In contrast to the large spin splitting in Co $3d$ states, the induced spin splitting in Pd $4d$ states at the interface exhibits a combination of up and down spin transitions that accompany the quadrupole-like states in Pd. In other words, the anisotropy of the orbital moments in Co is enhanced at the interface through the proximity with Pd and orbital moment anisotropy is not induced in Pd even in the facing layer on Co.

**Methods**

Two types of samples of ultrathin [Co/Pd]$_5$ multilayered structures were studied: Co (0.69 nm)/Pd (1.62 nm) that have PMA (Sample A), and Co (1.03 nm)/Pd (1.62 nm) that have an in-plane anisotropy (Sample B). They were grown on a Si(110) substrate at a substrate temperature of 150 °C, using the DC magnetron sputtering method with a Pd/Ta seed bilayer, as illustrated in Fig. 1. These samples were originally grown to study the photo-induced precession of magnetisation at a low energy excitation [18]. The 0.69-nm-thick Co and 1.62-nm-thick Pd in the Sample A correspond to four monolayers (MLs) and eight MLs thickness, respectively. Cross-sectional transmission electron microscopy images have shown ⟨111⟩-oriented layered metallurgical structures with a good interface abruptness between Co and Pd [18].

We performed XMCD experiments at the BL4B, UVSOR facility, Institute of Molecular Science (IMS) at a temperature of 10 K, and at the BL-7A in Photon Factory, high-energy accelerator organization (KEK) at room temperature in order to check the XMCD signals. We employed the total-electron-yield mode that directly detects the sample drain currents. Before the XMCD measurements, sample surfaces were sputtered by Ar-ions at a voltage of 1 kV, in order to remove the oxygen contamination. A magnetic field was applied along the direction of the incident polarised soft X-rays, using the setup reported in Refs. [35] and [36]. We reversed the magnetic field directions in order to provide right- and left-circular X-ray configurations for a fixed circular polarisation of the incident X-ray beam. Magnetic fields of ±1.2 T were applied during the measurements, sufficient to saturate the magnetisation. Angle-dependent XMCD was performed by changing the angle between the incident beam and the direction of the sample surface normal, starting from the surface normal to 60°; these geometries are defined as normal incidence (NI) and oblique or grazing incidence (GI), respectively. For the NI configuration, where both photon helicity and magnetic field direction are normal to the surface, the X-ray absorption processes involve the normal direction components of the orbital angular momentum ($m_{orb}^{\perp}$), deduced from the magneto-optical sum rule. The GI configuration mainly allows the detection of the in-plane orbital momentum components ($m_{orb}^{\parallel}$). Therefore, $\Delta m_{orb}$ can be deduced form the twice of the measured $m_{orb}^{\perp} - m_{orb}^{\parallel}$ values.

In order to apply the magneto-optical sum rules, background subtractions are essential. In Co $L$-edge XAS, we assumed the integral-type background for both pre- and post-edges being zero. In Pd $M$-edge XAS, we also used the integral-type background for only $M$-edge main peaks appearing at around 530 and 560 eV and fitted main peaks by Gaussian functions, and ignore other peaks which does not contribute to the XMCD, which provides an ambiguity (±20 %) for estimating the spin and orbital magnetic moments.

We have performed first-principles calculations for the Co/Pd stacked structures shown in Fig. 4b, assuming similar Co/Pd(111) multilayers structures as those used in the XMCD experiments. We employed the Vienna *ab initio* simulation package (VASP) using the projector augmented wave (PAW) potential [37] and the spin-polarised generalised gradient approximation (GGA) proposed by Perdew, Becke, and Ernzerhof for calculations of the exchange and correlation energies [38]. The Co/Pd(111) multilayers are constructed using a super-cell that has eight atomic layers of fcc-Pd and $n$ atomic layers of fcc-Co ($n$ = 1, 4, 7, and 10). The in-plane lattice constant is fixed to that of fcc-Pd ($a_{\parallel}$ = 0.391 nm), while the inter-layer distances along the <111> axis are energetically fully optimised for both Pd and Co, as well as for the Pd/Co interface. The $E_{MCA}$ is determined using the force theorem, where the differences in the sum of energy eigenvalues between the magnetisation oriented along the in-plane [110] and out-of-plane [111] directions are calculated including the spin-orbit interaction. The calculation of $E_{MCA}$ using VASP-PAW is discussed in detail in Ref. [39]. We also confirmed the similar results using the self-consistent calculations. Positive $E_{MCA}$ values imply a perpendicular magnetic anisotropy. The wavenumber $k$-point integration is performed using a modified tetrahedron method with Bloechl corrections [40], and 22×22×4 $k$-points in the first Brillouin zone of each Co/Pd(111) supercell.

**Acknowledgement**

This work was partially supported by JSPS KAKENHI (Grant Nos. 16H06332 and 15H03562), Spintronics Research Network of Japan, and the Creation of Life Innovation Materials for Interdisciplinary and International Researcher Development Project of MEXT. The authors would like to thank Dr. N. Nishizawa for his coordination in providing samples and associated structural and magnetisation data. The authors acknowledge Profs. Y. Takagi and T. Yokoyama at the IMS for the technical support for the synchrotron radiation experiments. Parts of the synchrotron radiation experiments were performed under the approval of the Photon Factory Program Advisory Committee, KEK (No. 2015G090).

**Author Contributions**

J.O. and H.M. planned the study. H.M. prepared samples and characterized the structural and magnetic properties. J.O. performed the XMCD measurements and analyzed the data. Y.M.

performed the band structure calculations. J.O. wrote the manuscript using the inputs from all authors. All authors discussed the results.

**Notes**

The authors declare no competing interests.

Figure captions:

**Figure 1 | Stacked structures and magnetic field dependence of the Co $L_3$-edge XMCD for a photon energy of 778.0 eV, in the NI and GI configurations.** (a) Co (0.6 nm)/Pd (1.62 nm) stacked structure that exhibits PMA (Sample A). (b) Co (1.03 nm)/Pd (1.62 nm) stacked structure that exhibits an in-plane magnetic anisotropy (Sample B).

**Figure 2 | $L$-edge XAS and XMCD of the Co/Pd multilayers that exhibit PMA, for Samples A and B.** (a) XAS for the NI configuration. (b) XMCD performed at the NI and GI configurations. (c) Integrals of the XMCD. The insets in (b) show the expanded regions of the $L_3$-edge.

**Figure 3 | $M$-edge XAS and XMCD of the Co/Pd multilayers that exhibit PMA.** (a) XAS at the NI configuration in left panel and the GI configuration at right panel. (b) XMCDs acquired for the NI and GI configurations, respectively. The integrals of the XMCD are also plotted in the figures.

**Figure 4 | DFT calculations.** (a) MCA energy $E_{\mathrm{MCA}}$, shape magnetic energy $2\pi M_S^2$, and $E_{\mathrm{MCA}}-2\pi M_S^2$ as a function of the thickness of the Co layer ($t_{\mathrm{Co}}$) for Pd(8ML)/Co($t_{\mathrm{Co}}$), with $a_{\parallel} = 0.391$ nm. (b) A schematic of the Pd(8ML)/Co(4ML)(111) multilayer. (c) Layer-resolved orbital moment anisotropy of the Pd(8ML)/Co(4ML), using $a_{\parallel} = 0.391$ nm.

**Figure 5 | DFT calculations.** Bar graph of the second-order perturbative contribution of the spin-orbit interaction to the MCA energy at the interfacial atomic sites of (a) Co and (b) Pd for the Pd(8ML)/Co(4ML), using $a_\parallel = 0.391$ nm. Spin-resolved local density of states (LDOS) of the $d(xy, x^2-y^2)$ and $d(yz, zx)$ states for the interfacial (c) Co and (d) Pd sites, for the Pd(8ML)/Co(4ML), using $a_\parallel = 0.391$ nm. (e) A schematic of the electron hopping in Co and Pd at the interface. $3d(xy, x^2-y^2)$ orbitals in Co and both Pd $4d(xy, x^2-y^2)$ and $4d(yz, zx)$ orbitals are illustrated.

Table I, The spin and orbital magnetic moments and magnetic dipole terms for perpendicular and in-plane directions in Sample A. The values are in the units of $\mu_B$ and are compared with the estimations from the XMCD and the first-principles DFT calculations. Experimental error bars are estimated about 10 % in Co and 20 % in Pd for the applications of XMCD sum rules.

|  | Co | | Pd | |
| --- | --- | --- | --- | --- |
|  | XMCD | DFT | XMCD | DFT |
| $m_{\mathrm{spin}}^{\perp}$ | 1.82 | 1.87 | 0.25 | 0.31 |
| $m_{\mathrm{spin}}^{\parallel}$ | 1.81 | 1.87 | 0.24 | 0.31 |
| $7m_{\mathrm{T}}$ | 0.01 | - | 0.01 | - |
| $m_{\mathrm{orb}}^{\perp}$ | 0.14 | 0.128 | 0.02 | 0.032 |
| $m_{\mathrm{orb}}^{\parallel}$ | 0.11 | 0.096 | 0.02 | 0.033 |

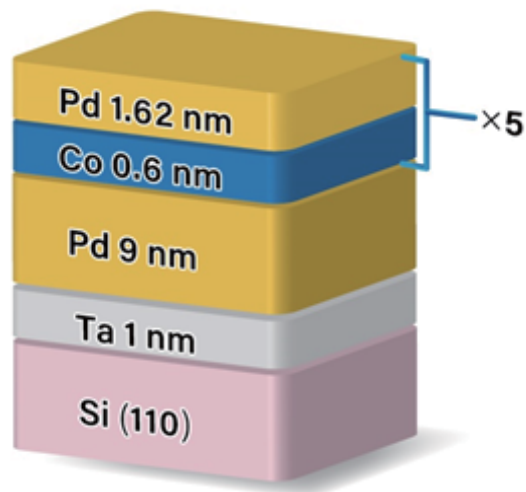
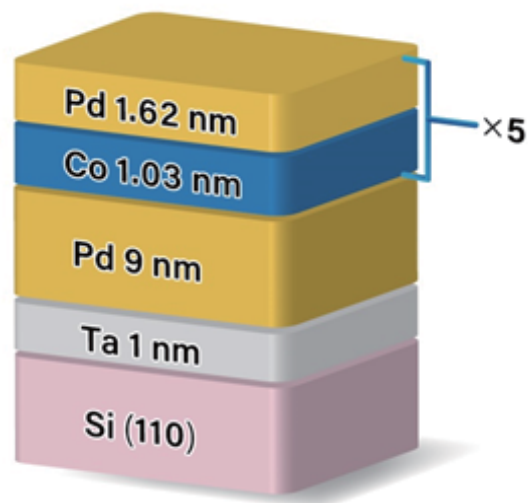
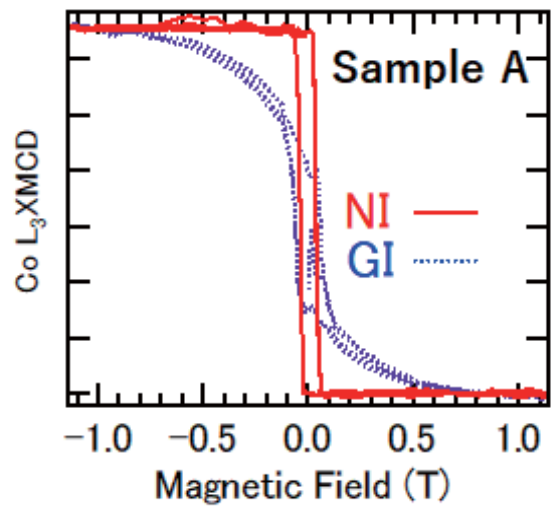
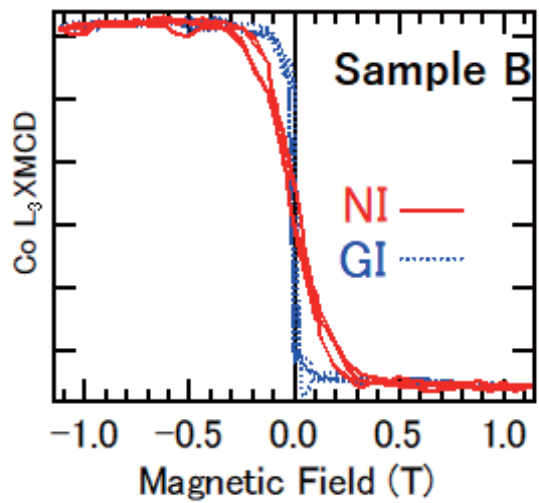

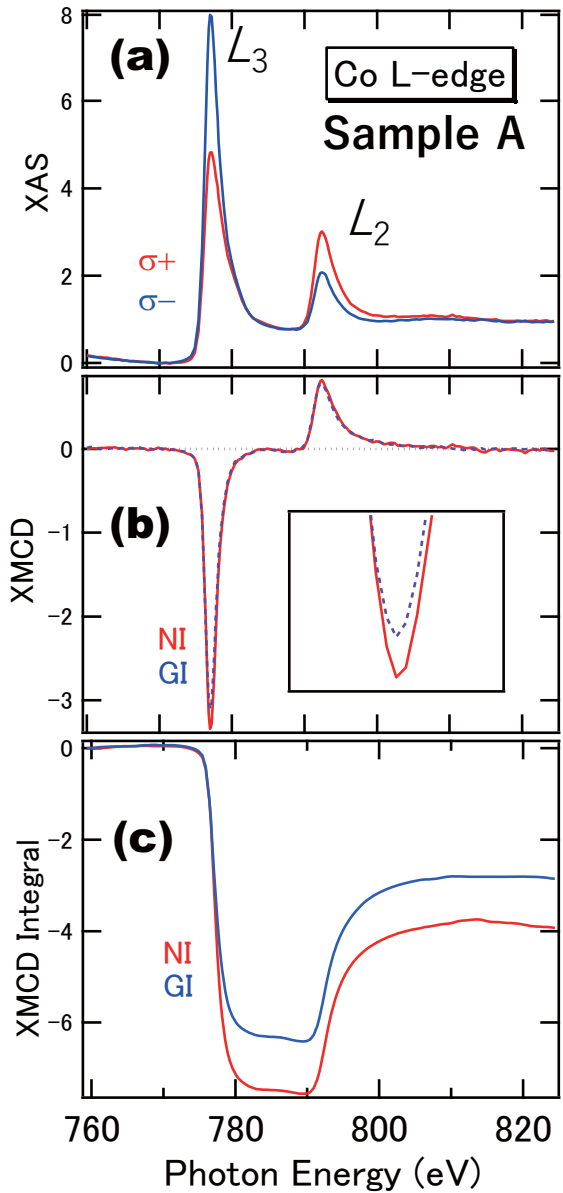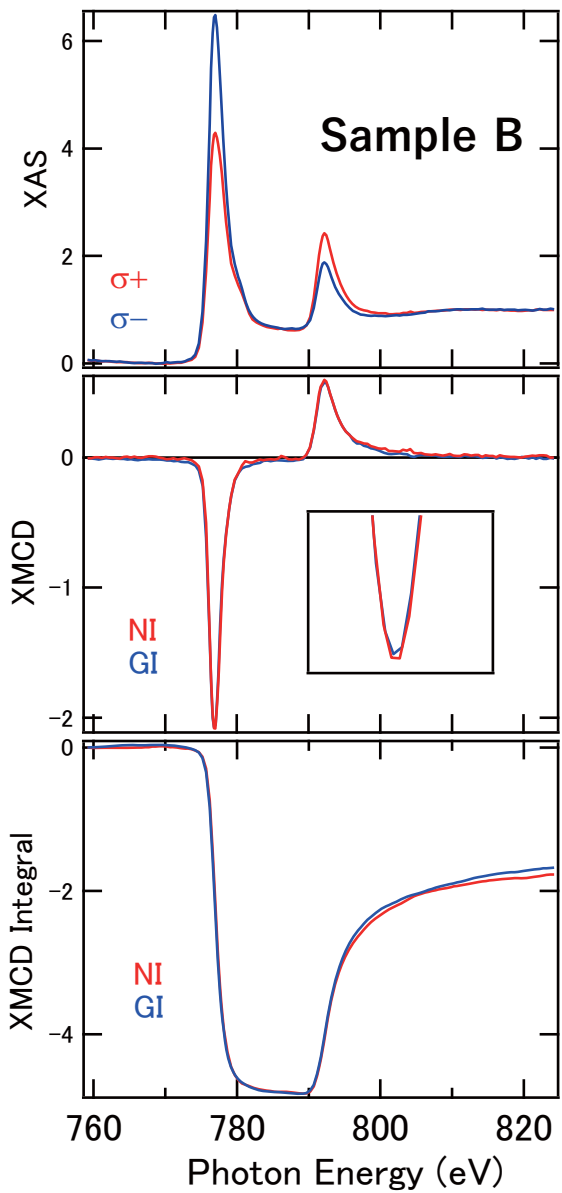

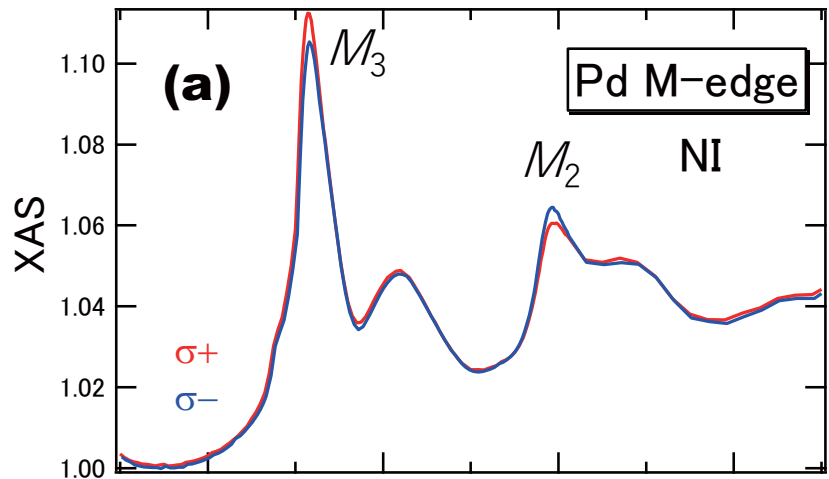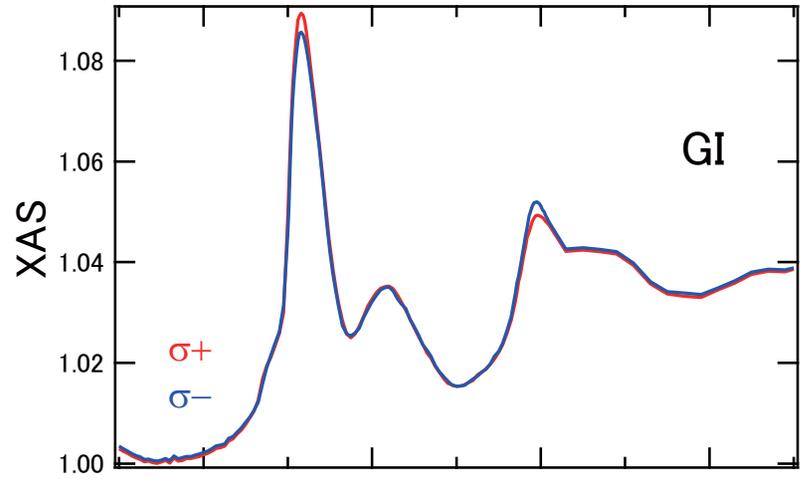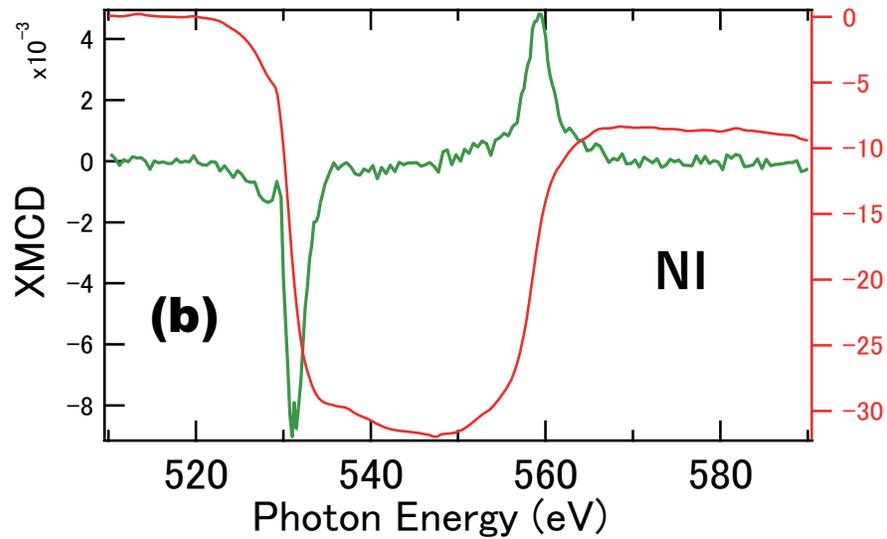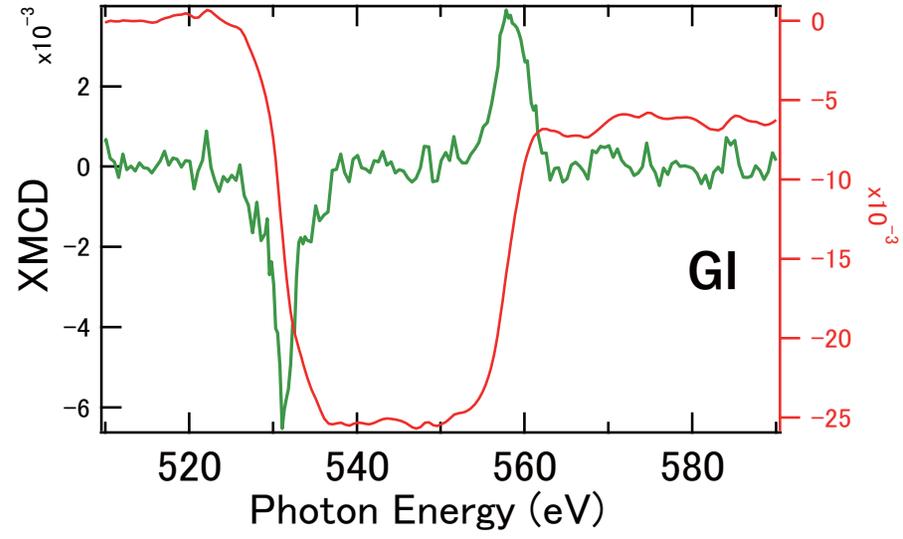

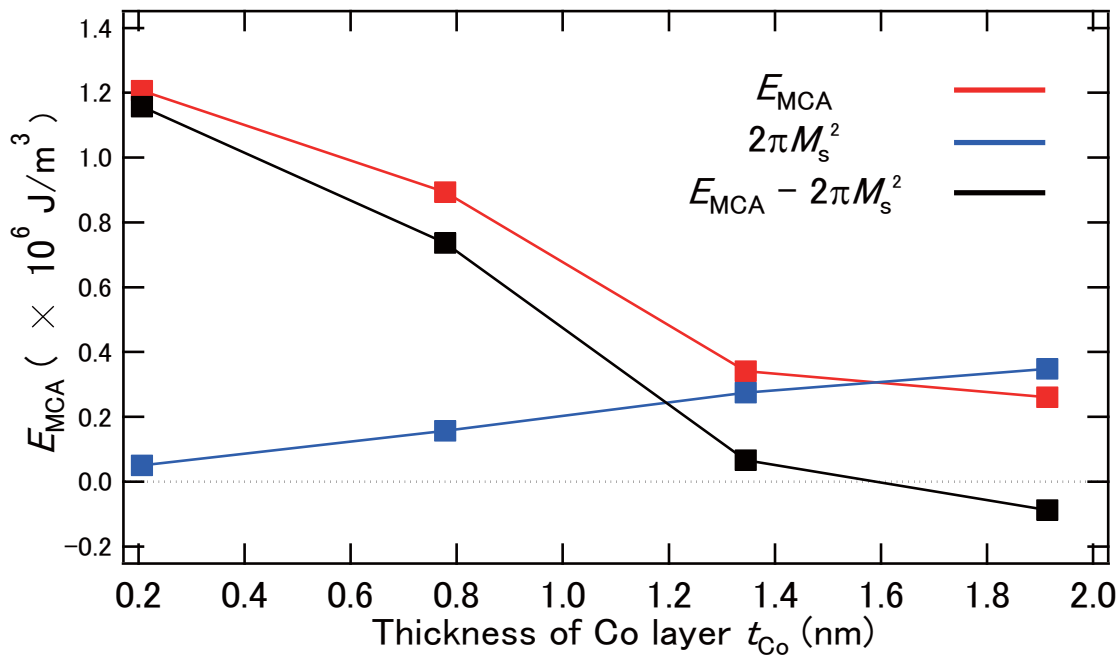

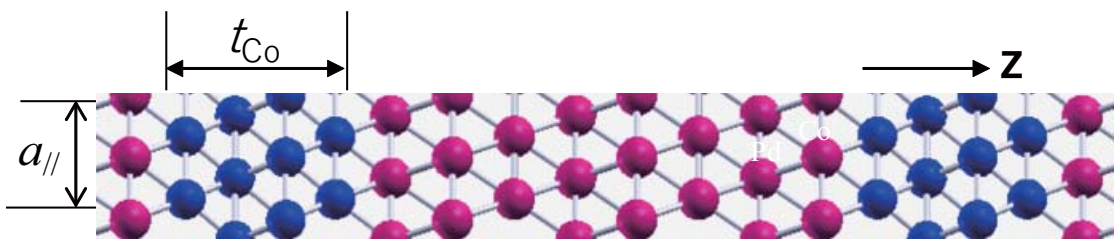

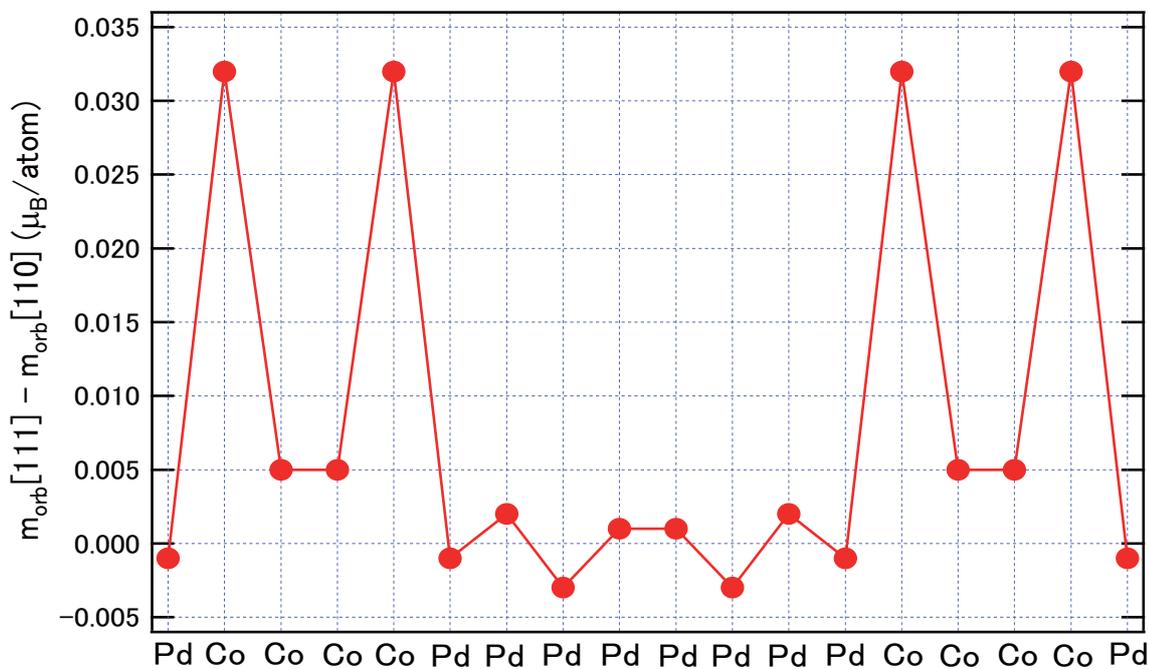

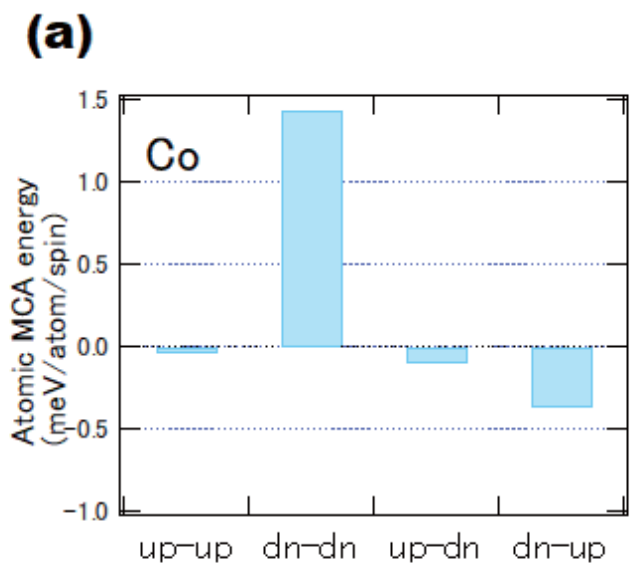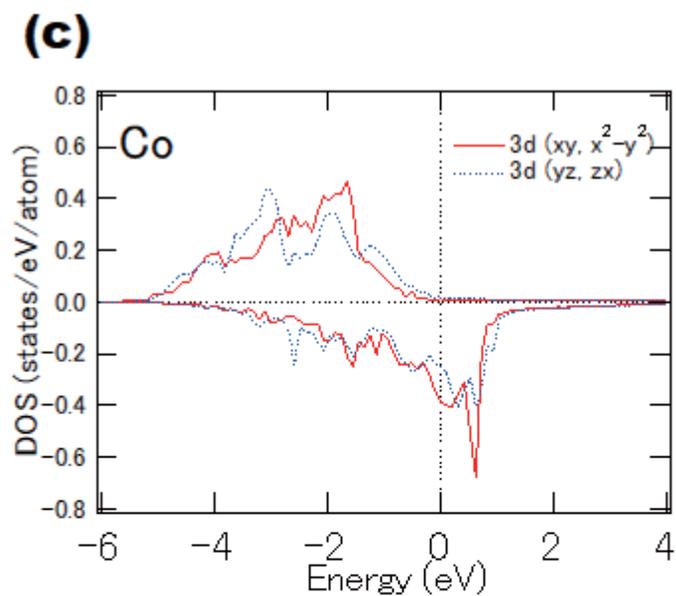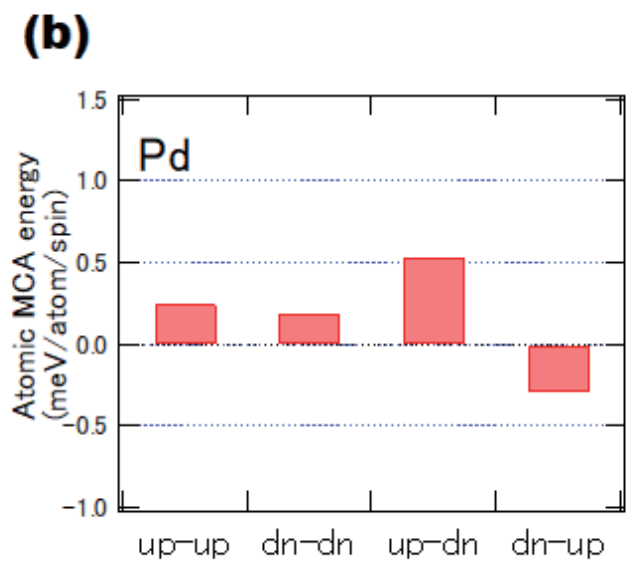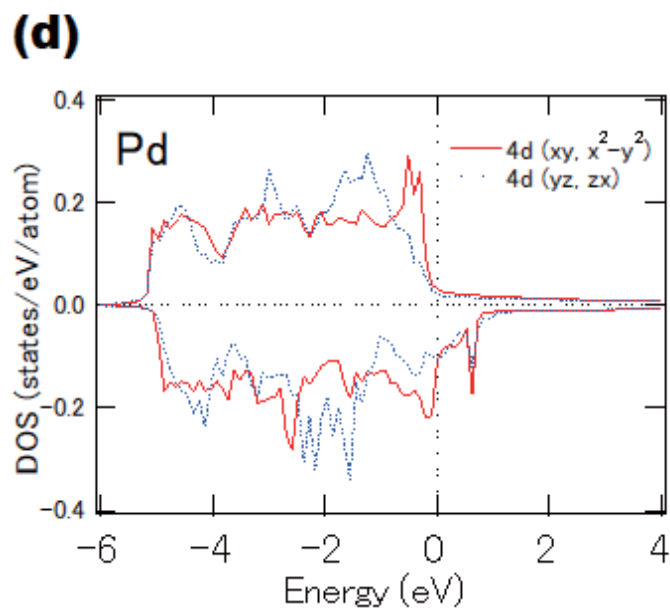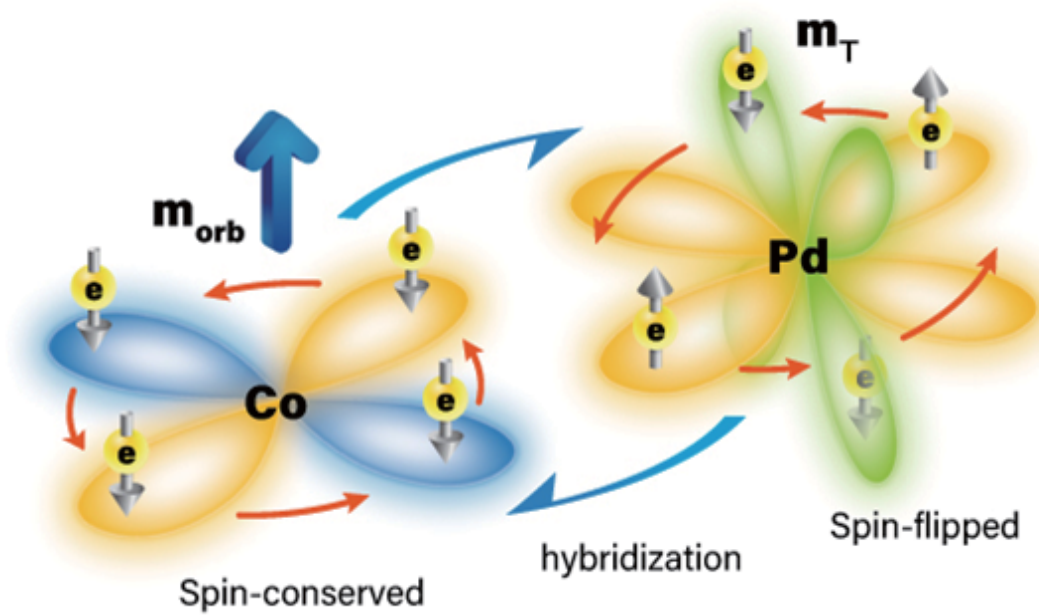